\documentclass[pra,aps]{revtex4}
\usepackage{graphicx}
\preprint{}
\begin{document} 
\title{A spherical perfect lens} 
\author{S. Anantha Ramakrishna} \affiliation{Department of Physics, Indian 
Institute of Technology, Kanpur 208016, India}
\author{J.B. Pendry} \affiliation{The Blackett Laboratory, Imperial College 
London, London SW2 2AZ, U.K.}
\date{\today}
\begin{abstract}
It has been recently proved that a slab of negative refractive index material
acts as a perfect lens in that it makes accessible the sub-wavelength 
image information contained in the evanescent modes of a source. Here we elaborate 
on perfect lens solutions to spherical shells of negative refractive 
material where magnification of the near-field images becomes possible. 
The negative refractive materials then need to be spatially dispersive with 
$\varepsilon(r) \sim 1/r$ and $\mu(r)\sim 1/r$. 
We concentrate on lens-like solutions for the extreme 
near-field limit. Then the conditions for the TM and TE polarized modes become independent of 
$\mu$ and $\varepsilon$ respectively.   
\end{abstract}
\maketitle
\section{Introduction}
The possibility of a perfect lens \cite{pendry_PRL00} whose resolution is not 
limited by the classical diffraction limit has been subject to intense debate 
by the scientific community during the past two years.
 This perfect lens could be realised by using a slab of material with 
$\varepsilon = \mu = -1$ where 
$\varepsilon$ is the dielectric constant and the $\mu$ is the magnetic 
permeability. Veselago had observed\cite{veselago} that such a 
material would have a negative refractive index of $n = -\sqrt{\varepsilon
\mu} = -1$ (the negative sign of the square root needs to be chosen by 
requirements of causality), and  a slab of such a material would act as a lens 
in that it would refocus incident rays from a point source on one side into 
a point on the other side of the slab (See Fig. 1). Due to the unavailability of 
materials with simultaneously negative $\varepsilon$ and $\mu$, negative refractive 
index remained an academic curiosity until recently when it became possible to 
fabricate structured meta-materials that have negative $\epsilon$ 
and $\mu$ \cite{smith00,smith01,eleftheriades}. Most of the negative refractive 
materials(NRM), so far, consist of interleaving arrays of thin metallic wires 
(that provide negative $\varepsilon$ \cite{pendry96}) and metallic split-ring 
resonators (that provide negative $\mu$ \cite{pendryIEEE}). 
Although some initial concerns were expressed 
\cite{garcia_OL} that the observed effects in these experiments were dominated by  
absorption, the recent experiments of Refs.\cite{parazzoli,houck,grbic,sridhar} have
confirmed that negative refractive materials are today's reality. 

\begin{figure}[b]
\includegraphics[width=8cm]{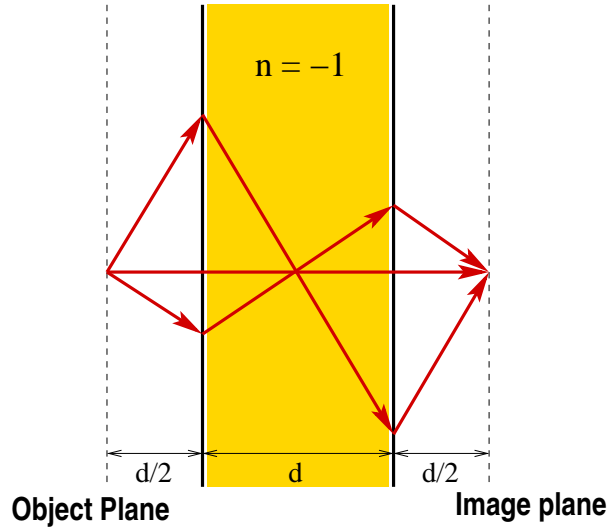}
\caption{Radiation from a point source on one side of a slab of material with i
$\varepsilon =-1$ and $\mu = -1$ is refocussed into a point on the other side. 
The rays representing propagating waves are bent on to the other side of the normal
at the interfaces due to the negative refractive index of the slab. }
\end{figure}
It was demonstrated by one of us that the NRM slab acts
a lens not only for the propagating waves (for which the ray analysis of Veselago
 is valid) but also for the evanescent near-field radiation \cite{pendry_PRL00}. 
This phenomenon of perfect lensing becomes possible due to the surface 
plasmon states \cite{raether} that reside on the
surfaces of the NRM slab which restore the amplitudes of the decaying evanescent
waves \cite{pendry_PRL00,drs_APL03,haldane,gomez_PRL,sar_JMO02,xsrao_PRB}. 
Indeed, it has been confirmed by numerical (FDTD) simulations that an incident pulse
is temporarily trapped at the interfaces for a considerable time \cite{foteinopolou}.
For a detailed description of the perfect slab lens, we refer the reader to Ref. 
\cite{pendry_PRL00,sar_JMO02,pendry_JPC02}.  The `perfectness' of the perfect lens
is limited only by the extent to which the constituent NRM are perfect with the
specified material parameters. Absorption in the NRM and deviations of the
material parameters from the resonant surface plasmon conditions of the perfect
lens causes significant degradation of the subwavelength resolution
possible \cite{drs_APL03,platzmann_APL,zye_PRB,fang_APL}.  We have
suggested some possible measures to ameliorate this degradation of the lens
resolution by stratifying the lens medium \cite{sar_JMO03} and introducing
optical gain into the system \cite{sar_PRBr03}.

The image formed by the NRM slab lens is identical to the object and hence there
is no magnification in the image. Lenses are mostly used to produce
magnified or demagnified images and the lack of any magnification is a great
restriction on the slab lens on which most of the attention in the literature
has been focussed. The slab lens is invariant in the transverse directions and
conserves the parallel component of the wave-vector. To cause magnification this
tranverse invariance will have to be broken and curved surfaces necessarily have 
to be involved. The perfect lens effect is dependent on the near-degeneracy of the
surface plasmon resonances to amplify the near-field, and curved surfaces in
general have completely different surface plasmon spectrum \cite{klimov}. 
It was recently pointed out by us that a family of near-field lenses (in 
the quasi-static approximation) in two-dimensions can be generated by a
conformal mapping of the slab lens \cite{pendry_JPC02}. Thus a cylinderical
annulus with dielectric constant $\varepsilon = -1$ was shown to have a
lens-like property of projecting in and out images of charge distributions. 
Similiarly in Ref. \cite{pendry_OE03} and \cite{pendry_JPC03}, it was shown how a 
general method of co-ordinate transformations could be used to map the perfect 
slab lens solution for the Maxwell's equations into a variety of situations 
including the cylinderical and spherical geometries respectively.

In this paper, we elaborate on the perfect lens solutions in the spherical
geometry and show that media with spatially dispersive dielectric constant
$\varepsilon(\vec{r})/sim 1/r$ and magnetic permeability $\mu(\vec{r})/sim 1/r$  
can be used to fabricate a spherical perfect lens that can magnify the near-field 
images as well. In section-2 of this paper, we will present
these perfect lens solutions of the Maxwell's equations for the spherical 
geometry. In section-3, we will examine the solutions in the extreme near-field
limit or the quasi-static limit which is useful when the lengthscales in the problem
are all much smaller than a wavelength. Then the requirements for TM and TE 
polarizations depend only on $\varepsilon \sim 1/r^2$ or $\mu \sim 1/r^2$ respectively. 
This is useful at frequencies where we are able to generate structures with only one of 
$\varepsilon$ or $\mu$ negative.  We will investigate the 
effects of dissipation in the NRM and point out the connections to the 1-D slab 
lens solutions. We will present our concluding remarks in Section-4.

\section{A perfect spherical lens}
Consider a spherically symmetric system shown in Fig. 2 consisting of a 
spherical shell of NRM with the dielectric constant $\varepsilon_{-}(r)$ and 
$\mu_-(r)$ imbedded in a positive refractive material with $\varepsilon_+(r)$ 
and $\mu_+(r)$. First of all we will find the general solutions to the field
equations with spatially inhomogeneous material parameters:
\begin{eqnarray}
\nabla & \times & \mathbf{E} = i \omega\mu_0 \mu(\mathbf{r}) \mathbf{H} ,
~~~~~~~~~~
\nabla \times \mathbf{H} = i \omega\varepsilon_0 \varepsilon
(\mathbf{r}) \mathbf{E} \\
\nabla & \cdot & \mathbf{D} = 0 , ~~~~~~~~~~ \nabla \cdot \mathbf{H} = 0, \\
\mathbf{D} &=& \varepsilon(\mathbf{r}) \mathbf{E}, ~~~~~~~~~~ \mathbf{B} =
\mu(\mathbf{r})
\mathbf{H}.
\end{eqnarray}

\begin{figure}[tbp]
\includegraphics[width=8cm]{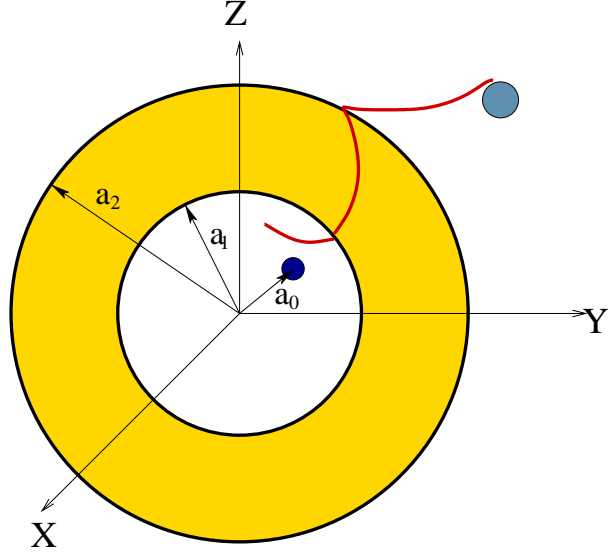}
\caption{A spherical shell with negative $\varepsilon_-(r) \sim -1/r$ and 
$\mu_-(r) \sim -1/r$ images a source located inside the shell into the external region. 
The media outside have postive refractive index, but  $\varepsilon_-(r)i\sim 1/r$ and
$\mu_-(r) \sim 1/r$. The amplification inside the spherical shell of the 
otherwise decaying field is schematically shown.} 
\end{figure}

Under these circumstances of spherical symmetry, it is sufficient to specify the
quantities $(\mathbf{r}\cdot\mathbf{E})$ and $(\mathbf{r}\cdot\mathbf{H})$ which
will constitute a full solution to the problem.
Let us now look at the TM polarised modes $\mathbf{r}\cdot\mathbf{H} = 0$,
implying that only the electric fields have a radial component $E_r$. 
Operating on Eqn. (1) by $\nabla$, we have
\begin{eqnarray}
\nabla\times\nabla\times\mathbf{E} &=& i \omega\mu_0 \nabla\times \left[ \mu(
\mathbf{r}) \mathbf{H} \right], \nonumber \\
&=& \frac{\omega^2}{c^2} \mu(\mathbf{r}) \varepsilon(\mathbf{r}) \mathbf{E} +
i \omega \frac{\nabla \mu(\mathbf{r})}{\mu(\mathbf{r})} \times
\nabla \times \mathbf{E}.
\end{eqnarray}
Using Eqns.(2) and (3)  we have
\begin{equation}
\nabla \cdot \mathbf{D} = \nabla \cdot \left[ \varepsilon(\mathbf{r})
\mathbf{E}\right] =  \nabla\varepsilon(\mathbf{r}) \cdot \mathbf{E} +
\varepsilon(\mathbf{r})~ \nabla \cdot \mathbf{E} = 0,
\end{equation}
 and if we assume $\varepsilon(\mathbf{r}) = \varepsilon(r)$ and $\mu(
\mathbf{r}) = \mu(r)$, we have
\begin{equation}
\nabla \cdot \mathbf{E} = -\frac{\varepsilon'(r)}{r \varepsilon(r)}
\mathbf{r} \cdot \mathbf{E}=  -\frac{\varepsilon'(r)}{r \varepsilon(r)} (rE_r).
\end{equation}
We note  the following identities for later use:
\begin{equation}
\nabla\times\nabla\times\mathbf{E} = \nabla (\nabla\cdot \mathbf{E}) -
\nabla^2 \mathbf{E},
\end{equation}
\begin{equation}
\nabla^2(\mathbf{r}\cdot \mathbf{E})  = \mathbf{r}\cdot \nabla^2\mathbf{E}
+ 2 \nabla\cdot \mathbf{E},
\end{equation}
and using Eqn. (6) we also note that
\begin{eqnarray}
\mathbf{r}\cdot\nabla (\nabla\cdot \mathbf{E}) &=&  \mathbf{r}\cdot\nabla
\left( -\frac{\varepsilon'(r)}{\varepsilon(r)} E_r \right), \nonumber \\
&=& -r \frac{\partial}{\partial r} \left( \frac{\varepsilon'(r)}{\varepsilon(r)}
E_r \right), \nonumber \\
&=& -\frac{\partial}{\partial r} \left( \frac{\varepsilon'(r)}{\varepsilon(r)}
(rE_r) \right) + \left( \frac{\varepsilon'(r)}{\varepsilon(r)} E_r \right).
\end{eqnarray}
We now take a dot product of $\mathbf{r}$  with Eqn. (4), and  use the
Eqns. (6),(7), (8) and (9) to get an equation for $(rE_r)$ as:
\begin{equation}
\nabla^2 (rE_r) + \frac{\partial}{\partial r} \left[ \frac{\varepsilon'(r)}
{ \varepsilon(r)}(rE_r) \right]  + \frac{\varepsilon'(r)}{r \varepsilon(r)}
(rE_r) + \varepsilon(r) \mu(r) \frac{\omega^2}{c^2} (rE_r) = 0 .
\end{equation}
This equation is separable and the spherical harmonics are a solution to the
angular part. Hence the solution is $(rE_r) = U(r) Y_{lm}(\theta,\phi)$ where
the radial part $U(r)$ satisfies
\begin{equation}
\frac{1}{r^2} \frac{\partial}{\partial r} \left( r^2 \frac{\partial U}{\partial
r} \right) - \frac{l(l+1)}{r^2} U + \frac{\partial}{\partial r} \left[
\frac{\varepsilon'(r)} { \varepsilon(r)}U \right] +
\frac{\varepsilon'(r)}{r
\varepsilon(r)} U + \varepsilon(r) \mu(r) \frac{\omega^2}{c^2} U = 0.
\end{equation}
If we choose $\varepsilon(r) = \alpha r^p$ and $\mu(r) = \beta r^q$, we can
have a solution $U(r) \sim r^n$ and we get
\begin{equation}
[n(n+1) - l(l+1) + p(n-1) + p] r^{n-2} +  \alpha\beta\omega^2/c^2 r^{p+q+n}
= 0,
\end{equation}
implying $p+q=-2$ and 
\begin{equation}
n_\pm=1/2 \left[ -(p+1) \pm \sqrt{(p+1)^2 + 4l(l+1)-4 \alpha\beta\omega^2/c^2} 
\right].
\end{equation}
Hence the general solution can be written as 
\begin{equation}
E_r(\mathbf{r}) = \sum_{l,m} \left[ n_+A_{lm} r^{n_+ -1} +
n_- B_{lm} r^{n_- -1}
 \right] Y_{lm}(\theta,\phi),
\end{equation}
and 
\begin{equation}
\mathbf{H}(\mathbf{r}) = \sum_{l,m}\left[ A_{lm} r^{n_+} + B_{lm} r^{n_-
} \right] \mathbf{X}_{lm}(\theta,\phi),
\end{equation}
where the vector spherical harmonic  
\begin{equation}
\mathbf{X}_{lm}(\theta,\phi) \equiv \mathbf{L} Y_{lm}(\theta,\phi) = \frac{1}{i}
(\mathbf{r}\times \nabla) Y_{lm}(\theta,\phi) . 
\end{equation} 
A similiar solution can be obtained for the TE modes with $\mathbf{r}\cdot
\mathbf{E} = 0$. 

Now assuming an arbitrary source at $r=a_0$, we can now write down the electric
fields of the TM modes in the different regions for the negative spherical shell 
of Fig. 2 as
\begin{eqnarray}
\mathbf{E}^{(1)}(\mathbf{r}) &=& \sum_{l,m}\left[ n_+ A_{lm}^{(1)} r^{n_+ -1}
+ n_- B_{lm}^{(1)} r^{n_- -1} \right] Y_{lm}(\theta,\phi), ~~~~~~~~~~
a_0<r<a_1,\\
\mathbf{E}^{(2)}(\mathbf{r}) &=& \sum_{l,m}\left[ n_+ A_{lm}^{(2)} r^{n_+ -1}
+ n_- B_{lm}^{(2)} r^{n_- -1} \right] Y_{lm}(\theta,\phi), ~~~~~~~~~~
a_1<r<a_2,\\
\mathbf{E}^{(3)}(\mathbf{r}) &=& \sum_{l,m}\left[ n_+ A_{lm}^{(3)} r^{n_+ -1}
+ n_- B_{lm}^{(3)} r^{n_- -1} \right] Y_{lm}(\theta,\phi), ~~~~~~~~~~
a_2<r<\infty,
\end{eqnarray}
and similiarly for the magnetic fields. Note that the $B_{lm}^{(1)}$ correspond
to the field components of the source located at $r=a_0$. For causal solutions
$A_{lm}^{(3)} = 0$. Now the tangential components of the magnetic fields and the
normal components of the displacement fields have to be continuous across the
interfaces.  Under the conditions $p = -1$, $q =-1$, $\varepsilon_+(a_1) =
-\varepsilon_-(a_1)$ and $\varepsilon_+(a_2) =  -\varepsilon_-(a_2)$, we have
\begin{eqnarray}
A_{lm}^{(1)} &=& 0,\\
A_{lm}^{(2)} &=&
\left(\frac{1}{a_1^2}\right)^{\sqrt{l(l+1)-\alpha\beta\omega^2/c^2}}
B_{lm}^{(1)}, \\
B_{lm}^{(2)} &=& 0,\\
B_{lm}^{(3)} &=&
\left(\frac{a_2^2}{a_1^2}\right)^{\sqrt{l(l+1)-\alpha\beta\omega^2/c^2}}
B_{lm}^{(1)}.
\end{eqnarray}
The lens-like property of the system becomes clear by writing the field outside
the spherical shell as
\begin{equation}
E_r^{(3)} = \frac{1}{r} \left[\frac{a_2^2}{a_1^2}
r\right]^{\sqrt{l(l+1)-\alpha\beta\omega^2/c^2}} B_{lm}^{(1)}
Y_{lm}(\theta,\phi).
\end{equation}
Hence apart from a scaling factor of $1/r$, the fields on the sphere
$r=a_3=(a_2^2/a_1^2)a_0$ are identical to the fields on the sphere $r =a_0$.
There is also a spatial magnification in the image by a factor of
$a_2^2/a_1^2$.

Let us note a couple of points about the above perfect lens solutions in the
spherical geometry. 
First, for $r>a_3$, i.e. points outside the image surface the fields appear as
if the source were located on the spherical image surface ($r = a_3$). However, this is
not true for points $a_2<r<a_3$ within the image surface.
Second, given that $\varepsilon_-(a_2) = -\varepsilon_+(a_2)$, we
have the perfect lens solutions if and only if $n_+ = - n_-$ which implies that
$p = -1$ in Eqn. (13). Although the  solutions given by Eqn. (14) 
occur in any
medium with $\varepsilon \mu \sim 1/r^2$, the perfect lens solutions only occurs
for $\varepsilon \sim\mu \sim 1/r$. Here we have written down the solutions for
the TM modes. The solutions for the TE modes can be similiarly obtained.

\section{The spherical near-field lens}
As it has been pointed out in the previous section, the `power' solutions 
are good for any $\varepsilon(r) \sim r^p$ and $\mu(r) \sim r^q$ such that $p+q
=-2$. However the perfect lens solutions for the Maxwell's equations  result 
only for the single case of $p = q =-1$. In the quasi-static limit of 
$\omega \rightarrow 0$ and $l \gg |p|,~|q|$, we can relax this condition. In particular, 
by setting $\varepsilon(r) \sim 1/r^2$ and $\mu(r)=$ constant, we can have a 
perfect lens for the TM modes alone. Similiarly, we can have a perfect lens for the
TE polarization by having $\mu \sim 1/r^2$ and $ \varepsilon =$ constant.

This extreme near-field limit is both important and valid for situations 
when all lengthscales in the problem are much smaller than a wavelength of
the radiation. This becomes useful at frequencies where we can only generate media
with either negative $\varepsilon$ and positive $\mu$, or, negative $\mu$ and 
positive $\varepsilon$. Examples are the silver slab lens at optical frequencies
\cite{pendry_PRL00}, the imeta-materials (Swiss rolls) 
used for MRI at radio-frequencies\cite{wiltshire}.
Particularly at radio- and microwave frequencies, we currently can practically 
engineer the required meta-materials with spatially dispersive characteristics 
at the corresponding length scales.  Further it 
also lifts the restriction that the system has to have a spatially 
dispersive material parameters even outside the spherical shell of NRM. In this section we 
will work in this extreme near field limit. Then it is sufficient to solve the Laplace 
equation and we present lens-like solutions to the Laplace equation
below. 
 
Consider the spherical shell in Fig.2 to be filled with a material 
with $\epsilon_2(r)\sim -C/r^2$ with the 
inner and outer regions filled with constant dielectrics of $\varepsilon_1$ and 
$\varepsilon_3$ respectively. Let $\mu =1 $ everywhere. 
Now place a charge $+q$ at the centre of the
concentric spheres and a charge $-q$ at a distance $a_0$ from the centre inside 
region-1. We will consider the z-axis to be along the dipole axis and make use 
of the azimuthal  symmetry here, although it is clear that our results do not 
depend on any such assumption of azimuthal symmetry. Thus all our charge and their
images will now lie along the Z-axis.

Now we will calculate the potentials in the three regions that satisfies the 
Laplace equation and the continuity conditions at the interfaces. The potential 
in region-1($r<a_1$)  can be calculated to be (using the azimuthal symmetry): 
\begin{equation} 
V_1(\mathbf{r}) = \frac{-q}{4\pi\epsilon_1} \sum_{l=1}^{\infty} \left[
A_{1l} r^l P_l(\cos\theta) + \frac{a_0^l}{r^{l+1}} P_{l}(\cos\theta) \right] 
\end{equation} 
Note that the second term in the above expansion arises  due to the dipole 
within the sphere.
It can be shown (see appendix-1), that the general form of the potential in 
region-2 ($a_1<r<a_2$), where the dielectric constant varies as $1/r^2$ is  
\begin{equation} 
V_2(\mathbf{r}) = \frac{-q}{4\pi\epsilon_0} \sum_{l=1}^{\infty} \left[
A_{2l} r^{(l+1)} P_l(\cos\theta) + \frac{B_{2l}}{r^{l}} P_{l}(\cos\theta)
\right]. 
\end{equation} 
In region-3($r>a_2$), the potential is given by:
\begin{equation} 
V_3(\mathbf{r}) = \frac{-q}{4\pi\epsilon_3} \sum_{l=1}^{\infty} \left[
\frac{B_{3l}}{r^{l+1}} P_{l}(\cos\theta) \right]. 
\end{equation} 
Now we must match the potentials at the interfaces at $r=a_1$ and $r=a_2$ (Put
$\epsilon_0 = 1$) to determine the $A$ and $B$ coefficients. The conditions of 
continuity of the potential and the normal component of $\vec{D}$ at the 
interfaces are 
\begin{eqnarray} 
V_1(a_1) = V_2(a_1) &,&~~~~~~~~~~V_2(a_2) = V_3(a_2), \\ 
\epsilon_1 \frac{\partial V_1(a_1)}{\partial r} = \epsilon_2 
\frac{\partial V_2(a_1)}{\partial r} &,& ~~~~~~~~~ \epsilon_2 \frac{\partial 
V_2(a_2)}{\partial r} = \epsilon_3 \frac{\partial V_3(a_2)}{\partial r} 
\end{eqnarray} 
We determine the coefficients from these conditions to be: 
\begin{eqnarray} 
A_{1l} &=&  \frac{(l+1)a_0^l \{[l\epsilon_2(a_2) - (l+1)
\epsilon_3][\epsilon_1+\epsilon_2(a_1)] 
- [\epsilon_2(a_2)+\epsilon_3]
[(l+1)\epsilon_1-l\epsilon_2(a_1)] \frac{a_2^{2l+1}}{a_1^{2l+1}}\} }
{l(l+1)[\epsilon_1+\epsilon_2(a_1)][\epsilon_2(a_2)+\epsilon_3] a_2^{2l+1}+ 
[l\epsilon_1 - (l+1) \epsilon_2(a_1)][l \epsilon_2(a_2) - (l+1)\epsilon_3] 
a_1^{2l+1}},\\ 
A_{2l} &=& \frac{(2l+1)[l \epsilon_2(a_2) - (l+1)\epsilon_3]a_0^l a_1^{-1}} 
{l(l+1)[\epsilon_1+\epsilon_2(a_1)][\epsilon_2(a_2)+\epsilon_3] a_2^{2l+1}+
[l\epsilon_1 - (l+1) \epsilon_2(a_1)][l \epsilon_2(a_2) - (l+1)\epsilon_3] 
a_1^{2l+1}},  \\
B_{2l} &=& \frac{(2l+1)(l+1) [\epsilon_2(a_2) + \epsilon_3] a_2^l a_0^l 
a_1^{-(l+2)}} 
{l(l+1)[\epsilon_1+\epsilon_2(a_1)][\epsilon_2(a_2)+\epsilon_3] a_2^{2l+1}+ 
[l\epsilon_1 - (l+1) \epsilon_2(a_1)][l \epsilon_2(a_2) - (l+1)\epsilon_3] 
a_1^{2l+1}}, \\ 
B_{3l} &=& \frac{(2l+1)^2 \epsilon_3\epsilon_2(a_2) a_0^l a_2^{2(l+1)} a_1^{-1}} 
{l(l+1)[\epsilon_1+\epsilon_2(a_1)][\epsilon_2(a_2)+\epsilon_3] a_2^{2l+1}+ 
[l\epsilon_1 - (l+1) \epsilon_2(a_1)][l \epsilon_2(a_2) - (l+1)\epsilon_3] 
a_1^{2l+1}}.  
\end{eqnarray} 
Under the perfect lens conditions  
\begin{equation} 
\varepsilon_2(a_1) = -\varepsilon_1, ~~~~\mathrm{and}~~~~ 
\varepsilon_2(a_2) = -\varepsilon_3, 
\end{equation} 
we have: 
\begin{eqnarray} 
A_{1l} &=& 0, \\ 
A_{2l} &=& \frac{1}{\varepsilon_1} \frac{a_0^l}{a_1^{2(l+1)}},\\ 
B_{2l} &=& 0,\\
B_{3l} &=& \frac{\varepsilon_3}{\varepsilon_1} \left(\frac{a_2}{a_1}
\right)^{2(l+1)} 
 a_0^l. 
\end{eqnarray} 
Hence the potential outside the spherical shell for $r>a_2$ is 
\begin{equation} 
V_3(\vec{r}) = \frac{-q}{4\pi\varepsilon_3} \sum_{l=1}^\infty 
\frac{\varepsilon_3}{\varepsilon_1} \left(\frac{a_2}{a_1} \right) ^{2(l+1)} 
\frac{a_0^l}{r^{l+1}}, 
\end{equation} 
which is the potential of a dipole with the positive charge at the origin and 
the negative charge at $a_3$, where 
\begin{equation} 
a_3 = \left( \frac{a_2}{a_1}\right) ^2 a_0, 
\end{equation} 
and of strength  
\begin{equation} 
q_2 = \frac{\varepsilon_3}{\varepsilon_1} \left(\frac{a_2}{a_1}\right)^2 q = q, 
\end{equation} 
as $\varepsilon_3/\varepsilon_1 = (a_1/a_2)^2$.
Thus, on one side of the image (the region $r > a_3$) the fields of a point charge located 
at $a_3$ are reproduced. However it should be pointed  out that 
 there is no physical charge in the image location and, the fields on the 
other side of the image (i.e. in the region $a_2< r < a_3$) do not converge to the fields 
of the object and cannot do so in the absence of a charge in the image. 
Further there is no change in the  strength of the charge either. There is a 
magnification in the image formed by a factor of $(a_2/a_1)^2$.

Now let us consider the case of a point source placed at $a_3$ in the outer 
region. Again assuming the z-axis to pass through $a_3$, we can 
write the potentials in the three regions as  
\begin{eqnarray}
V_1(\mathbf{r})&=& \frac{+q}{4\pi\varepsilon_1}\sum_{l=0}^{\infty}
A_{1l}r^l P_l(\cos\theta)~~~~~~~~~~~~~~ \forall ~~~~~~r<a_1, \\ 
V_2(\mathbf{r})&=& \frac{+q}{4\pi}\sum_{l=0}^{\infty} \left[A_{2l}r^{l+1}+
\frac{B_{2l}}{r^l}\right] P_l(\cos\theta)~~~~~~~~~~~~~~ \forall ~~~~~~a_1<r<a_2, \\ 
V_3(\mathbf{r})&=& \frac{+q}{4\pi\varepsilon_3}\sum_{l=0}^{\infty}
\left[ \frac{r^l}{a_3^{l+1}} + \frac{B_{3l}}{r^{l+1}}\right] P_l(\cos\theta),
~~~~~~~~~~~~~~\forall ~~~~~~a_2<r<a_3, 
\end{eqnarray}
where the first term in $V_3(\vec{r})$ comes from the point source at $a_3$. Now
applying the conditions of continuity at the interfaces, we can similiarly
obtain for the coefficients as before.
In the limiting case of $\varepsilon_2(a_1) = -\varepsilon_1$ and
$\varepsilon_2(a_2) = -\varepsilon_3$, we have
\begin{eqnarray}
A_{1l} &=& \frac{\varepsilon_1}{\varepsilon_3} \left(\frac{a_2}{a_1}\right)^{2l}
\frac{1}{a_3^{l+1}}, \\
A_{2l} &=& 0, \\
B_{2l} &=& \frac{1}{\varepsilon_3} \frac{a_2^{2l}}{a_3^{2l+1}},  \\
B_{3l} &=& 0.
\end{eqnarray}
Hence the potential inside the inner sphere is
\begin{equation}
V_1(\mathbf{r}) = \frac{q}{4\pi\varepsilon_1} \sum_{l=0}^{\infty}
\frac{\varepsilon_1}{\varepsilon_3} \left(\frac{a_2}{a_1}\right)^{2l}
\frac{r^l}{a_3^{l+1}} P_l(\cos\theta),
\end{equation}
i.e., that of a point charge of strength $q_1 = (\varepsilon_1/\varepsilon_3)
(a_1/a_2)^2 = q$ at $a_0 = a_3 (a_1/a_2)^2$. 
As before, for the inner region of $r < a_0$, the system behaves as if there were
a single charge of strength $q$ located at $r=a_0$.  Thus the shell
has a lens-like action. We note that there is a demagnification of $(a_1/a_2)^2$
in this case.

\subsection{Similiarities to the 1-D slab lens}
	Let us point out the similiarities to the planar slab lens.
In both cases, the electromagnetic field grows in amplitude across the negative
medium when the perfect lens conditions are satisfied at the interfaces: as an
exponential($\exp [+k_x z]$) in the planar lens and as a power of the radial
distance $r^l$ in the spherical lens. The decaying solution away from the source
is absent in the negative medium in both cases.  Further, when  the perfect lens
conditions are matched at both the interfaces, there is no reflected wave in 
both the planar slab as well as the spherical lens: i.e. the impedance matching 
is perfect as well. In addition this mapping preserves the strength of the 
charge. 

	The key differences, however, are the different dielectric constants on either 
sides of the spherical shell of the negative medium. This is a direct 
consequence of the spatial $1/r^2$ dependence of the negative dielectric 
constant which relates the two positive dielectric constants to be  $\epsilon_1 
= (a_1/a_2)^2 \epsilon_3$. But this need not be a particular restriction as we
can use the ideas of the asymmetric lens to terminate the different positive 
media at some radii beyond\cite{sar_JMO02}. The net result is that the image can now be
magnified (or demagnified) when the image of the charge (source) 
 is projected out of (or into) the spherical shell, which is true 
in the 2D cylinderical lens as well \cite{pendry_JPC02}.

\subsection{Possibility of the asymmetric lens}
	In the case of a planar slab, it was possible to have the perfect lens effect 
by satisfying the required conditions at any one interface - not necessarily at 
both interfaces\cite{sar_JMO02}. Particularly , in the limit of very large 
parallel wave vectors the lensing is indeed perfect, although the image 
intensity differed from the source by a constant factor. Similiarly let us now 
investigate the effects of having the perfect lens conditions in the case of the 
spherical lens at only one of the interfaces. 
 
Let us consider first, the case of projecting out the image of a point source from inside 
the spherical shell to outside and enable the perfect lens conditions only at 
the outer interface $\epsilon_2(r=a_2) = -\epsilon_3$ and have an arbitrary 
$\epsilon_1$. Now the $A$ and $B$ coefficients come out to be
\begin{eqnarray}
A_{1l} &=& \frac{a_0^l}{a_1^{2l+1}} \frac{(l+1)
[\varepsilon_1+\varepsilon_2(a_1)]}
{l\varepsilon_1 - (l+1)\varepsilon_2(a_1)} \\
A_{2l} &=& \frac{(2l+1) a_0^l}{l\varepsilon_1 - 
(l+1)\varepsilon_2(a_1) a_1^{2l+2}} \\
B_{2l} &=& 0 \\
B_{3l} &=& \frac{(2l+1) \varepsilon_3 a_0^{l}}{l
\varepsilon_1(a_1/a_2)^2 +
(l+1)\varepsilon_3} \left( \frac{a_2}{a_1} \right)^{2l}
\end{eqnarray}
Only the growing solution within the negative spherical shell remains. The 
coefficient of the decaying solution ($B_{2l}$) remains strictly zero. Thus, 
amplification of the decaying field at least is possible in this case as well. 
But there is a  finite reflectivity in this case. However, the solution outside for $r>a_2$ 
is not the exact image field of the point source as the coefficient $B_{3l}$ 
has an extra dependence on $l$ through the dependence on the dielectric constants. 
Moreover, the process does not preserve the strength of the charge due to the 
different dielectric constants involved. 
This should be compared to the solution of the planar asymmetric slab lens 
where, at least  in the electrostastic limit, the system behaved as a perfect 
lens. In this case, the system behaves as a spherical asymmetric perfect lens only in the 
limit of large $l \rightarrow \infty$. The solution outside the spherical shell 
is the same when we meet the perfect lens condition on the inner interface -- just 
as in the case of the planar slab lens.  However, the  reflection coefficient is 
again non-zero, but different to the earlier case. In either case, the fields 
are largest at the interface where one meets the perfect lens conditions or the
interface on which the surface plasmons are excited. 

\subsection{Effects of dissipation} 
Media with negative real part of the dielectric constant are absorptive (as all
metals are), and hence we can write the dielectric constant $\varepsilon(r) =
C/r^2 + i \varepsilon_i(r)$ (Note that $\varepsilon_i \sim 1/r^2$ as well for us
to be able to write the solution in the following form). Consider the first
case of projecting out the image of a dipole located within the spherical shell 
 where the potential  outside the shell is given by 
Equation (27) and $B_{3l}$ is given by Equation (33).
When we have a dissipative negative medium and have the perfect lens
conditions at the interfaces on the real parts of the dielectric constant alone,
$\varepsilon_2(a_1) = -\varepsilon_1 + i \varepsilon_i (a_1)$ and
$\varepsilon_2(a_2) = -\varepsilon_3+i\varepsilon_i (a_2)$. In parallel with the
case of the planar lens, we note that the denominator of $B_{3l}$ consists of
two terms, one containing a power of the (smaller) radius $a_1$ and the other
containing a power of the (larger) radius $a_2$. Crucially the amplification of
the evanescent fields depends on the possibility that the smaller power
dominates by making the coefficient of
the larger term as close as possible to zero. The presence of the imaginary part
of the dielectric constant would not allow the coefficient to be zero and the
image restoration is good only as long as the term containing $a_1$ dominates in
the denominator of $B_{3l}$, i.e.,
\begin{equation}
l(l+1) \varepsilon_i(a_1)\varepsilon_i(a_2) a_2^{2l+1}
\ll [(2l+1)\varepsilon_1 -i(l+1)
\varepsilon_i(a_1)][-(2l+1)\varepsilon_3+i\varepsilon_i(a_2)]a_1^{2l+1}
\end{equation}
Hence a useful estimate of the extent of image resolution can be obtained by
noting the multipole $l$ for  which the two terms in the denominator are
approximately equal\cite{drs_APL03}. We obtain for this value:
\begin{equation}
l_{\mathrm{max}} \simeq \frac{ \ln \left\{ 3\varepsilon_1 \varepsilon_3 /
[\varepsilon_i(a_1)\varepsilon_i(a_2)] \right\}}{2 \ln \left( a_2/a_1
\right) }.
\end{equation}
Higher order multipoles are essentially unresolved in the image. 
We can similiarly  obtain the same criterion by considering the second case of 
tranferring the image of a charge located outside the spherical shell into the inner region.
Again, we can consider the effects of deviating 
from the perfect lens conditions on the real part of the dielectric constant as 
well and obtain a similiar limit for the image resolution due to those 
deviations.

\section{Conclusions}

In conclusion, we have presented a spherical perfect lens which enables 
magnification of the near-field images. The perfect lens solution requires media with 
$\varepsilon(r) \sim 1/r$ and $\mu(r) \sim 1/r$ 
and the conditions $\varepsilon_-(a_{1,2}) = -\varepsilon_+(a_{1,2})$ and $\mu_-(a_{1,2}) =
-\mu_+(a_{1,2})$ at the interfaces of the spherical shell of the NRM.
We have shown that in the quasi-static limit of small frequencies ($\omega
\rightarrow 0$) and high-order multipoles $l \gg |p|$, this condition can be 
relaxed and the two polarizations (TE and TM modes) decouple. Thus a shell with 
negative dielectric constant $\varepsilon_-(r) \sim -1/r^2$ with $\mu = 
\mathrm{constant}$ can act as a near-field lens for the TM polarization while 
$\mu_-(r) \sim -1/r^2$ with $\varepsilon_-(r) = \mathrm{constant}$ acts as a 
near-field lens for the TE modes.  We have shown that 
dissipation in the lens material, however, prevents good resolution of higher order
multipoles. Thus while the near-field lenses work best for the higher order 
multipoles, dissipation cuts-off the higher order multipoles. Further  the 
spherical lens works in the asymmetric mode only in the limit of high order 
multipoles. Thus one has to find an intermediate regime where dissipation does 
not wipe out the near-field image information and yet the meta-materials work. 
This is the design challenge involving these near-field lenses.

\section*{Appendix: Solution of the Laplace equation in a spatially varying  
medium}
We have to solve the Maxwell's equations in material media 
\begin{equation}
\nabla \cdot\mathbf{D} = 0,~~~~\Rightarrow~~~~~ \nabla(\varepsilon
\mathbf{E}) = 0
\end{equation}
Using $\mathbf{E} = -\nabla V$, where $V(\mathbf{r})$ is the electrostatic
potential we have:
\begin{equation}
\varepsilon(\mathbf{r}) \nabla^2 V + \nabla \varepsilon(\mathbf{r})
\cdot \mathbf{\nabla}V = 0
\end{equation}
If $\varepsilon(\mathbf{r})$ has only a radial dependence
(as in our case $\sim 1/r^2$), then $\nabla \varepsilon(\mathbf{r}) =
\hat{r} (\partial \varepsilon / \partial r)$ and we can separate the 
solution as $V(\mathbf{r}) = \frac{U(r)}{r} Y_{lm}(\theta,\phi)$ where the $Y_{lm}$
is the spherical harmonic and the radial part $U(r)$ satisfies:
\begin{equation}
\varepsilon(r) \frac{ d^2 U}{d r^2} - \frac{l(l+1)}{r^2} U +\frac{d
\varepsilon}{dr} \left[ \frac{d U}{d r} - \frac{U}{r} \right] = 0
\end{equation}
To have a solution as a single power of $r$, the only choices possible
for the dielectric constant are either $\varepsilon = C$, a constant
-- the usual case , or
$\varepsilon = C/r^2$. In the latter case the solution is $U(r) \sim r^{l+2}$
or $U(r) \sim r^{-(l-1)}$. The full solution can then be written as
\begin{equation}
V(\mathbf{r}) = \sum_{l=0}^{\infty} \left[ A_{lm} r^{l+1} + B_{lm}r^{-l}
\right] Y_{lm}(\theta,\phi).
\end{equation}

\end{document}